\documentclass[aps,pra,twocolumn]{revtex4}

\usepackage{graphicx}
\usepackage{amsmath}
\usepackage{amssymb}

\usepackage{bm}
\usepackage{hyperref}
\usepackage{array}

\newcommand{\p}{\partial}

\newcommand{\ez}{\bm{\hat{e}}_3}

\newcommand{\emn}{\epsilon_{\mu\nu}}

\newcommand{\eln}{\epsilon_{\lambda\nu}}
\newcommand{\half}{\frac{1}{2}}

\newcommand{\lex}{\ell_{\rm ex}}

\newcommand{\Ea}{E_{\rm a}}
\newcommand{\hext}{h_{\rm ext}}
\newcommand{\bhext}{\bm{h}_{\rm ext}}

\newcommand{\skyrmion}{n}
\newcommand{\Skyrmion}{\mathcal{N}}

\newcommand{\storque}{\beta}
\newcommand{\polarity}{\lambda}
\newcommand{\winding}{\kappa}
\newcommand{\VAsize}{d_{\rm VA}}
\newcommand{\Je}{J_{\rm e}}
\newcommand{\Jp}{J_{\rm p}}

\newcommand{\bz}{{\bar{z}}}

\newcommand{\Omegax}{X}
\newcommand{\bOmegax}{\overline{X}}

\begin{document}

\title{Magnetization oscillations by vortex-antivortex dipoles}
\author{Stavros Komineas}
\affiliation{Department of Applied Mathematics, University of Crete, 71409 Heraklion, Crete, Greece}

\begin{abstract}
A vortex-antivortex dipole can be generated due to current with in-plane spin-polarization, 
flowing into a magnetic element,
which then behaves as a spin transfer oscillator.
Its dynamics is analyzed using the Landau-Lifshitz equation including 
a Slonczewski spin-torque term.
We establish that the vortex dipole is set in steady state rotational motion due to
the interaction between the vortices, 
while an external in-plane magnetic field can tune the frequency of rotation.
The rotational motion is linked to the nonzero skyrmion number of the dipole.
The spin-torque acts to stabilize the vortex dipole at a definite vortex-antivortex separation distance.
In contrast to a free vortex dipole, the rotating pair under spin-polarized current 
is an attractor of the motion, therefore a stable state.
Three types of vortex-antivortex pairs are obtained as we vary the external field and spin-torque strength.
We give a guide for the frequency of rotation based on analytical relations. 
\end{abstract}

\maketitle


\section{Introduction}
\label{sec:intro}

The injection of a dc spin-polarised current through a magnetic element can induce
magnetisation oscillations and thus turn a nanoelement into a spin-transfer-torque oscillator
\cite{RussekRippardCecil_2010}.
This property can be exploited for the design of, possibly, the smallest available frequency generators.
The magnetisation configuration which is set in periodic motion may itself be 
a nonlinear magnetic excitation, such as a magnetic vortex.

Magnetic vortices are typically seen to be created and annihilated in pairs in numerical simulations
due to spin-polarised current \cite{BerkovGorn_JAP2006,BerkovGorn_JPD2008,BerkovGorn_PRB2009}.
In the experiments in Ref.~\cite{FinocchioOzatay_PRB2008} spin-polarized current 
was injected through an aperture into an elliptic-shaped magnetic element.
Accompanying simulations showed a spontaneously nucleated vortex-antivortex (VA) pair,
where vortex and antivortex have opposite polarities, in rotational motion.

The dynamics of vortex pairs has been previously studied
in the context of the conservative Landau-Lifshitz (LL) equation
\cite{Pokrovskii_JETP1985,PapanicolaouSpathis_NL1999,Komineas_PRL2007}.
Experimental and numerical results show that the creation
and stabilisation of such VA pairs can occur under spin-polarised current,
that is, in a non-conservative system excitable by an external probe. 
The robustness of VA pairs in the excitable system is our motivation in order to study
those in detail theoretically and numerically.

In the experiments in Ref.~\cite{FinocchioOzatay_PRB2008} the polarisation of the spin current is
in the plane of the film, 
while an external field, which is optionally applied, is also in-plane.
This obviously breaks the symmetry for the magnetisation in the plane and a preferential
direction arises.
However, the resulting vortex rotation in the plane surprisingly appears to restore the symmetry.

The presence of many parameters (strength of spin-torque, external field, damping, etc)
which contribute to the observed phenomena make the system complicated and
allow little space for an intuitive understanding.
This is stressed by a number of experiments conducted, under different conditions, which
have measured signals attributed to nonuniform or nonlinear magnetic states
\cite{RippardPufallKaka_PRL2004,RippardPufallKaka_PRB2004,PufallRippardSchneider_PRB2007}.

We will give a systematic theoretical and numerical study for vortex-antivortex dipoles in a current with
in-plane spin-polarisation.
This follows ideas already presented in Ref.~\cite{Komineas_EPL2012}.
We give a resolution of the mechanism for VA pair rotation 
which is necessary in order to understand the features of this system.
We also give analytical formulae for the frequency of rotation and for the magnetisation configuration
which will be used as a guide to explore the parameter space of this system.

The outline of the paper is as follows.
In Sec.~\ref{sec:llgs} we give the equation of motion.
In Sec.~\ref{sec:steadystate} we give a general description of vortex-antivortex pairs
and some analytical results.
In Sec.~\ref{sec:simulations} we give the results of numerical simulations.
In Sec.~\ref{sec:uniform} we present rotating VA pairs for the special case of uniform spin-torque.
Sec.~\ref{sec:conclusions} contains our concluding remarks.
Some calculations and detailed analytical results are given in Appendices.

\section{The Landau-Lifshitz-Gilbert-Slonczewski equation}
\label{sec:llgs}

The standard dynamics of the magnetisation vector $\bm{m}=(m_1,m_2,m_3)$ is given by 
the Landau-Lifshitz-Gilbert (LLG) equation.
A spin-polarised current is assumed to flow through an ultrathin film inducing excitation of the
magnetisation which can be described by an additional, so called,  Slonczewski spin-torque term in the LLG equation \cite{slonczewski_JMMM96} (see \cite{BerkovMiltat_JMMM2007} for a review).
The model which will be used throughout this paper is 
the Landau-Lifshitz-Gilbert-Slonczewski (LLGS) equation, in rationalised form,
\begin{align}  \label{eq:llgs}
&\dot{\bm{m}} = 
 -\bm{m}\times ( \alpha_1\bm{f} -  \alpha_2\storque\,\bm{p} ) 
 -\bm{m}\times \left[\bm{m}\times (\alpha_2\, \bm{f} + \alpha_1\storque\,\bm{p}) \right]  \notag \\
 & \bm{f} \equiv \Delta\bm{m} - m_3\ez + \bhext.  
\end{align}
Damping is accounted for by the coefficients $\alpha_1=1/(1+\alpha^2),\; \alpha_2=\alpha/(1+\alpha^2)$,
where $\alpha$ is the Gilbert dissipation constant.
We have taken into account three terms in the effective field $\bm{f}$: the exchange interaction,
an easy-plane anisotropy term perpendicular to the third magnetisation direction $\ez=(0,0,1)$, and an external field $\bhext$.
The spin-polarisation of the current is along the direction $\bm{p}$ which is taken to be a constant vector.
We further assume that the coefficient of the spin-torque term $\storque$ is a constant, thus we
confine ourselves to studying a simple form of the spin-torque term.

We consider Eq.~\eqref{eq:llgs} as a two-dimensional model, that is the magnetisation
is a function of two spatial variables and time, $\bm{m}=\bm{m}(x,y,t)$.
This is assumed to be the limit of an ultrathin film, and
the easy-plane anisotropy term is considered mainly as an approximation to the magnetostatic energy.

The magnetisation $\bm{m}$ and the field $\bhext$ are measured
in units of the saturation magnetisation $M_s$, so $\bm{m}^2=1$.
The units of length and time are, respectively,
\begin{equation}  \label{eq:lex_tau0}
\lex \equiv \sqrt{\frac{2A}{\mu_0 M_s^2}},\qquad \tau_0 = (\gamma \mu_0  M_s)^{-1},
\end{equation}
where $A$ is the exchange constant, $\gamma$ is the gyromagnetic ratio, and $\lex$ is the exchange length.
The spin-torque parameter $\storque$ is defined by
\begin{equation}  \label{eq:storque}
\storque = \frac{\Je}{\Jp},\qquad
\Jp = \frac{\mu_0 M_s^2 |e| d_f}{\hbar},
\end{equation}
where $\Je$ is the current density and $d_f$ is the thickness of the film.
For permalloy we have $\lex \approx 7\,{\rm nm}$, and
$\tau_0 \approx 7\,{\rm psec}$ which corresponds 
to a frequency $f_0=1/(2\pi\,\tau_0) \approx 23\,{\rm GHz}$,
and typically $\Jp \approx 3\times 10^{12}\,{\rm A/m}$.

\section{Steady-state rotating dipoles}
\label{sec:steadystate}

\subsection{Vortex-antivortex pair}

Vortices are magnetisation configurations where the magnetisation vector rotates a full $2\pi$ angle
when a circle is traced around a point called the vortex center. 
Vortices are thus characterised by a {\it winding number} $\winding=\pm 1$.
The positive value corresponds to the vortex typically observed in magnetic elements
which minimizes the magnetostatic energy.
A vortex with a negative winding number is called an antivortex.
Easy-plane anisotropy supports vortices with their {\it polarity}
along the third axis taking the two values $\polarity=\pm 1$.

A configuration of a vortex-antivortex (VA) pair has been
conjectured to form during dynamical processes in experiments, 
and the process has been demonstrated numerically.
We will study only the case where the vortex and the antivortex have opposite polarities.
Such a VA dipole has a nonzero skyrmion number.
The latter is defined as
\begin{equation}  \label{eq:skyrmion}
\Skyrmion = \frac{1}{4\pi} \int \skyrmion\,d^2x,\qquad \skyrmion = \half \emn (\p_\nu \bm{m} \times \p_\mu \bm{m})\cdot \bm{m},
\end{equation}
where $\skyrmion$ is a local topological density and $\emn$ is the totally antisymmetric tensor with $\mu,\nu=1,2$ \cite{rajaraman}.
A VA dipole has $\Skyrmion=\pm 1$.
We conventionally take the vortex with negative polarity and the antivortex with positive polarity, 
so the pair has $\Skyrmion=1$.
Changing the polarities of both vortices would change the sign of $\Skyrmion$.

Due to the interaction between the two vortices the pair cannot be static.
The vortices rotate clockwise or anticlockwise depending on the sign of $\Skyrmion$.
The situation is clear within the conservative LL equation, i.e., Eq.~\eqref{eq:llgs} 
for $\alpha=0, \storque=0$ and no external field $\bhext=0$.
The pair is pinned at the position where it is created and their dynamics is rotational \cite{Pokrovskii_JETP1985}.
This can be linked to the nonzero skyrmion number \cite{Komineas_PRL2007} and
 the vortices rotate clockwise or anticlockwise depending on the sign of $\Skyrmion$.
It is seen numerically that their motion can be close to a steady state rotation, 
however, this appears to be unstable.
If we add dissipation $\alpha \neq 0$ to the model then the vortices go on a spiralling orbit towards each other.

The rotational motion is stable in the case of a VA dipole under the influence
of a spin-polarised current flowing through a thin magnetic element,
as shown in experiments and simulations \cite{FinocchioOzatay_PRB2008,Komineas_EPL2012}.
The extensive numerical simulations in Sec.~\ref{sec:simulations} show that
the rotational motion is a stable steady state for a range of parameter values.
We stress that both the current polarisation \eqref{eq:p} and the external field \eqref{eq:hext}
will be considered to be in-plane, along the $x$-axis.
In such a case one would, in general, expect precession of the magnetisation around the $x$ axis.
So, the observed rotation is not easily interpreted as a straightforward consequence 
of the precessional dynamics of the Landau-Lifshitz equation.

A simple ansatz for a VA dipole can be written through the stereographic variable
discussed in Appendix \ref{app:X}.
The form \eqref{eq:VAx} represents a VA configuration which is an exact solution of the LL equation
when only the exchange interaction is considered.
The roles of the external magnetic field and the spin-torque term manifest themselves 
explicitly in this case \cite{Komineas_EPL2012}. 
The spin-torque acts to stabilise the radius of rotation while the external
field gives the angular frequency of rotation.
The rotation frequency is inversely proportional to the skyrmion number,
which manifests that it is crucial that $\Skyrmion \neq 0$ for these results to be valid.

\subsection{Virial relation}

We consider a uniform ferromagnetic state $\bm{m}_0=(1,0,0)$.
Such a magnetisation orientation may be due to the shape of the film as, the elliptic shape 
of the sample in Ref.~\cite{FinocchioOzatay_PRB2008}, or it may be imposed 
through the application of an external magnetic field.

The polarisation $\bm{p}$ of the current is assumed
\begin{equation}  \label{eq:p}
\storque\, \bm{p} = \storque\,  (1,0,0),\qquad  \storque < 0.
\end{equation}
that is, it tends to induce magnetisation antiparallel to $\bm{m}_0$.
Following experimental setups we mostly assume that the current flows through a nano-aperture
with a diameter of tens of nanometers. However, we also study the case of uniform spin current
in Sec.~\ref{sec:uniform}.
Whenever an external magnetic field is present this is considered uniform and has the form
\begin{equation}  \label{eq:hext}
\bhext = (\hext,0,0),\qquad \hext > 0,
\end{equation}
that is, it favors the uniform in-plane magnetisation.

Based on the evidence from numerics we conjecture the existence of a magnetic configuration
in steady-state rotation, in the sense of Eq.~\eqref{eq:rigid_rotation}.
The approximation of a steady-state allows for the derivation of explicit results.
Appendix \ref{app:virial} details a virial relation of Derrick type in Eq.~\eqref{eq:derrick} 
which is exact in the case of steady-state rotation.
According to the results of numerical simulations a simplified form of the Derrick relation \eqref{eq:derrick}
holds to a very good approximation for most of the VA dipoles presented in this paper.
This gives the angular velocity of rotation as
\begin{equation}  \label{eq:derrick_approx}
\omega \doteq -\left( \frac{\Ea}{\ell} + \hext \frac{\mu_1}{\ell} \right),
\end{equation}
where the symbol $\doteq$ indicates an approximation.
The quantity
\begin{equation}  \label{eq:angular_momentum}
 \ell = \half \int \rho^2\,\skyrmion\,d^2x,\qquad \rho^2 \equiv x_1^2+x_2^2,
\end{equation}
defined through the topological density $\skyrmion$, is identified with the angular momentum \cite{Papanicolaou91}.
For a VA pair with well-separated vortex and antivortex at a distance $\VAsize$,
we have $\ell \approx \Skyrmion (\pi/2) \VAsize^2$.
We will actually define the vortex-antivortex distance (for $\Skyrmion=1$) as
\begin{equation}  \label{eq:VAsize}
\VAsize \equiv \sqrt{\frac{2\ell}{\pi}}.
\end{equation}
The quantity
\begin{equation}  \label{eq:wa}
\Ea = \half \int (m_3)^2\,d^2x
\end{equation}
is the anisotropy energy, which takes the value $\Ea=\pi/2$ for a single isolated vortex \cite{Komineas_NL1998}.
We have also defined
\begin{equation}  \label{eq:magnetisation1}
 \mu_1 = -\frac{1}{2} \int x_\mu \p_\mu m_1\, d^2x = \int (1-m_1)\,d^2x,
\end{equation}
where the last equation derives from a partial integration assuming vanishing boundary terms,
and the final quantity gives the total magnetisation (spin reversals) in the $x$ axis.

Relation \eqref{eq:derrick_approx} establishes that the angular frequency of a rotating VA pair
has two separate contributions. The first term on the right hand side
is due to the interaction between the vortex and the antivortex.
This decreases as the distance between the vortices increases ($\ell$ increases).
The second term on the right hand side of \eqref{eq:derrick_approx} is due to the external field
and it includes a factor which depends on the details of the magnetic configuration.
It is crucial for both factors that $\Skyrmion  \neq 0$, since otherwise the quantity $\ell$
could be vanishing and change completely the meaning of the Derrick relation.

\section{Numerical Simulations}
\label{sec:simulations}

\subsection{Magnetisation configurations}

We have performed a series of numerical simulations based on Eq.~\eqref{eq:llgs}.
Following experiments we assume that the spin current is injected in the nanoelement through an aperture.
We model the flow of the current through a nano-aperture
by assuming a spin-torque parameter $\storque \neq 0$ 
in a disc with diameter $d_a \approx 40 {\rm nm}$ while we set $\storque = 0$
outside it.
For spin polarisation as in Eq.~\eqref{eq:p}
numerical simulations \cite{FinocchioOzatay_PRB2008} have shown that the spin-torque causes the
magnetisation to switch at the central area of the element and finally induces the generation 
of a VA dipole in rotational motion around the center of the spin current.
Mechanisms for the generation of a VA dipole are presented by numerical
simulations in Ref.~\cite{BerkovGorn_PRB2009}.
In the present paper we {\it assume} the existence of a vortex dipole and study its subsequent dynamics.

As the number of parameters in the present problem is large
we will not explore all possibilities but we will rather fix the parameters
\begin{equation}  \label{eq:parameters}
\alpha = 0.02,\qquad d_a=6\lex,
\end{equation}
for the dissipation and the aperture diameter respectively.
Numerical simulations performed with similar values for $\alpha$ have given
quantitatively similar results.
Other values for $d_a$ (e.g., $d_a=4\lex$ and $12\lex$)
did not give qualitatively different results. 
However, more extensive simulations would be needed in order to explore the dependence of results
on $d_a$.

We evolve an initial ansatz for a VA pair configuration in time according to the 
LLGS Eq.~\eqref{eq:llgs} and this typically converges to a steady-state rotating VA pair.
As the same result is obtained for any reasonable initial condition we conclude 
that these are stable dynamical magnetic configurations.
We study the features of our system by systematically varying
the parameters which can be tuned experimentally: the externally applied field $\hext$
and the strength of the spin current $\storque$.
Simulations are performed in two-dimensions using stretched coordinates for the infinite plane.

\begin{figure}[t]
   \centering
   \includegraphics[width=2.5in]{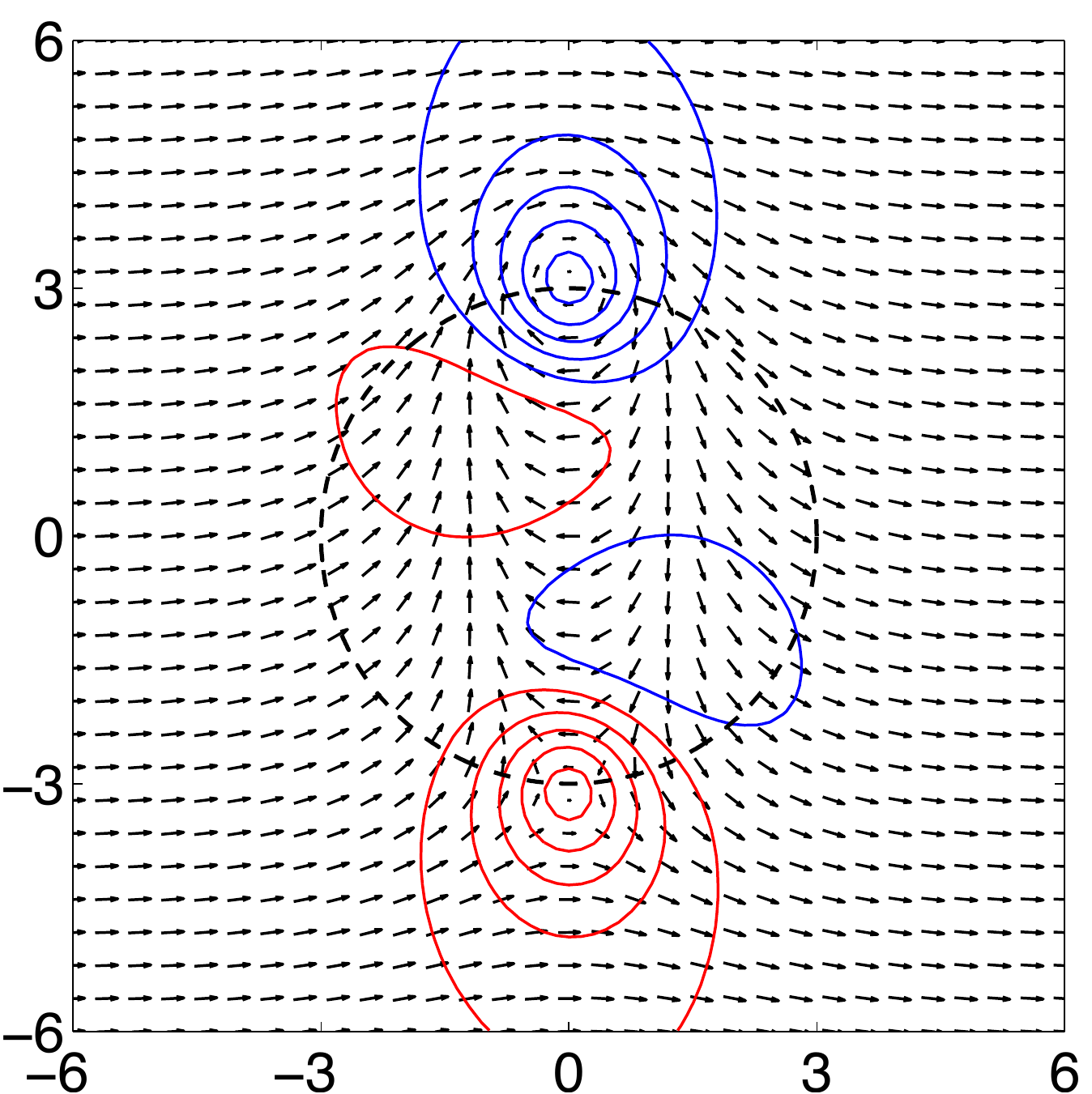} 
   \caption{Magnetisation configuration of type I (long VA pair) for vortex-antivortex pair 
   in steady-state rotation around a nano-aperture (delimited by a dashed line). 
   The vector plot shows $(m_1,m_2)$ and the contour plot is for the perpendicular magnetisation component. We plot contour levels $m_3 = \pm 0.1, \pm 0.3, \pm 0.5, \pm 0.7, \pm 0.9$ (red: up, blue: down).
   Parameters used are 
   $\storque=-0.10,\; \hext=0.40$ and \eqref{eq:parameters}. 
   The angular frequency of rotation is $\omega=0.255$.}
   \label{fig:VAlong}
\end{figure}

We start by choosing a typical value for the current $\storque=-0.1$.
We then vary the external field and find a series of
rotating VA pairs.
Varying $\hext$ from the value $\hext=0$ up to $\hext=0.44$ we find configurations similar in form to that
shown in Fig.~\ref{fig:VAlong} for the specific value $\hext=0.4$.
We label these VA pairs as type I (or {\it long} VA pairs)
and note that the vortex and antivortex are well separated.

\begin{figure}[t]
   \centering
   \includegraphics[width=2.5in]{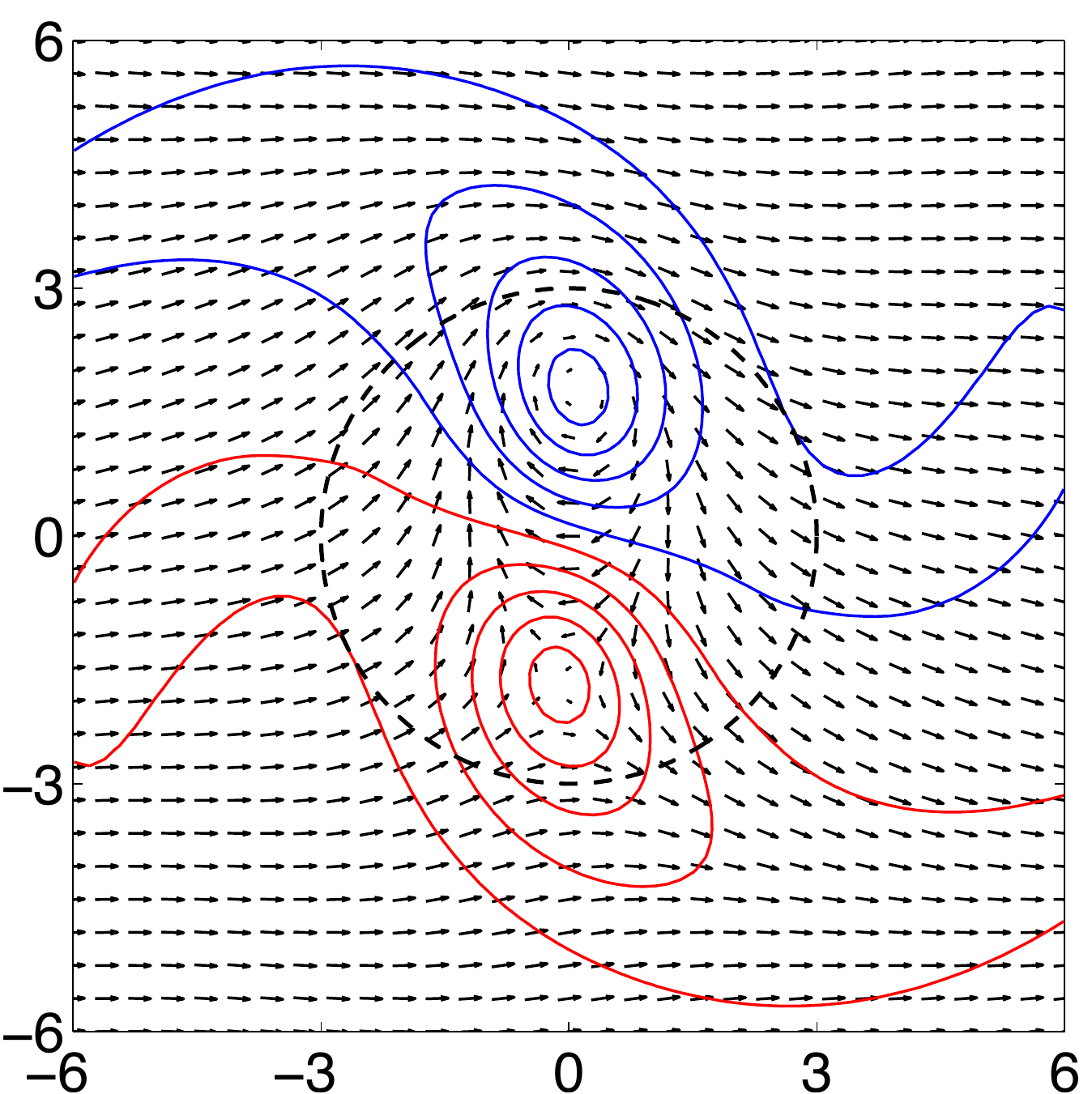} 
   \caption{Magnetisation configuration of type II (short VA pair) in steady-state rotation.
   We plot as in Fig.~\ref{fig:VAlong}.
   Parameters used are $\storque=-0.10,\; \hext=0.40$ (the same as in Fig.~\ref{fig:VAlong}). 
   The angular frequency of rotation is $\omega=0.539$.}
   \label{fig:VAshort}
\end{figure}

Varying $\hext$ from larger to smaller values we find configurations similar in form to that
shown in Fig.~\ref{fig:VAshort} for the value $\hext=0.4$.
We label these VA pairs as type II (or {\it short} VA pairs).
Note that the two VA pairs shown in Figs.~\ref{fig:VAlong}, \ref{fig:VAshort} are obtained
for the same parameter values, but they are different and they have significantly
different rotation frequencies.

\begin{figure}[t]
   \centering
   \includegraphics[width=2.5in]{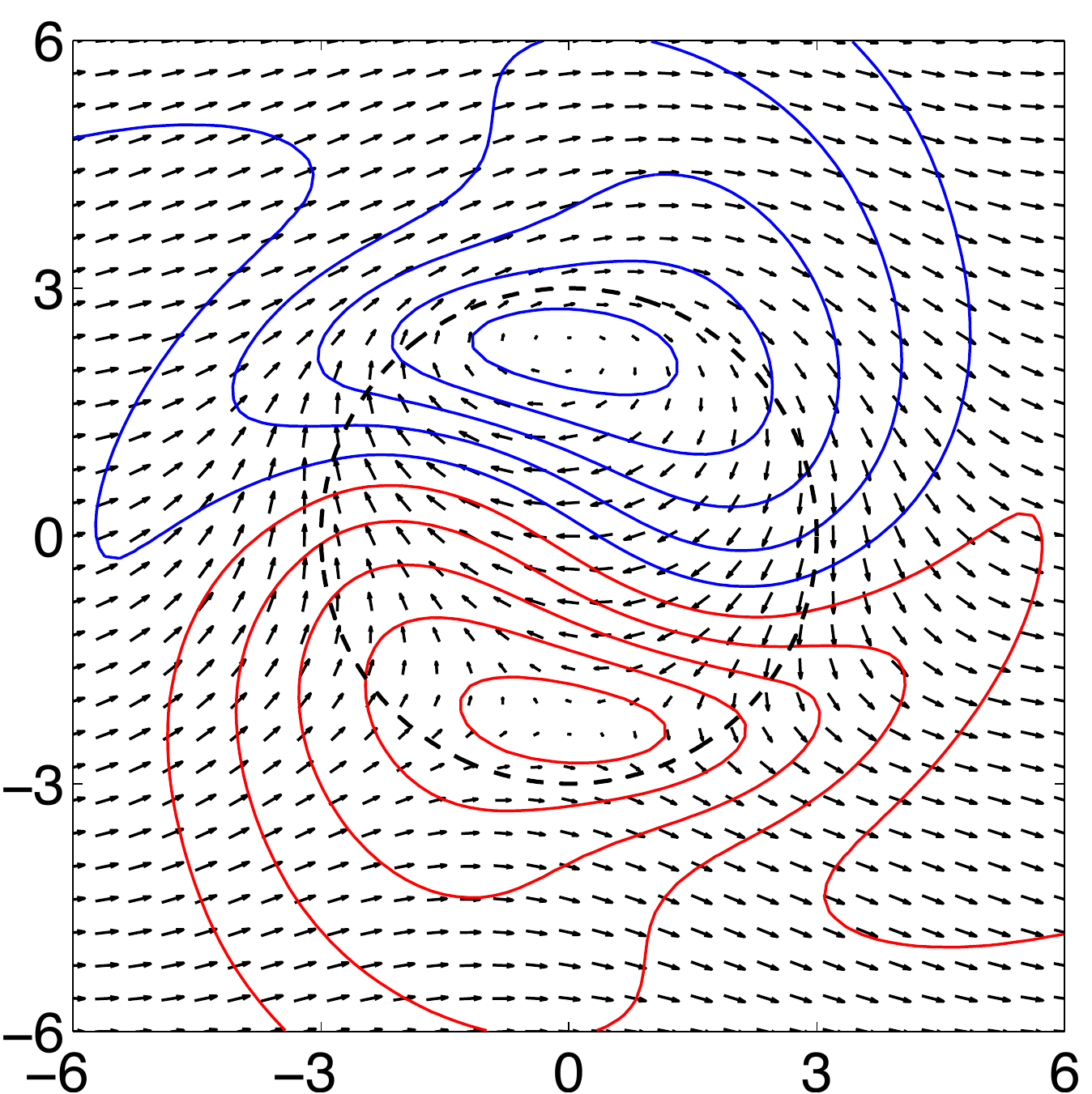} 
   \caption{Magnetisation configuration of type III (wide VA pair) in steady-state rotation.
    We plot as in Fig.~\ref{fig:VAlong}.
   Parameters used are $\storque=-0.20,\; \hext=0.60$. 
   The angular frequency of rotation is $\omega=0.820$.}
   \label{fig:VAwide}
\end{figure}

Increasing the spin current to $\storque=-0.2$ and choosing large external field values
we find still one more kind of VA pairs, shown in Fig.~\ref{fig:VAwide} for the
value $\hext=0.6$.
We label these VA pairs of type III (or {\it wide} VA pairs).

\begin{figure}[t]
   \centering
   \includegraphics[width=2.5in]{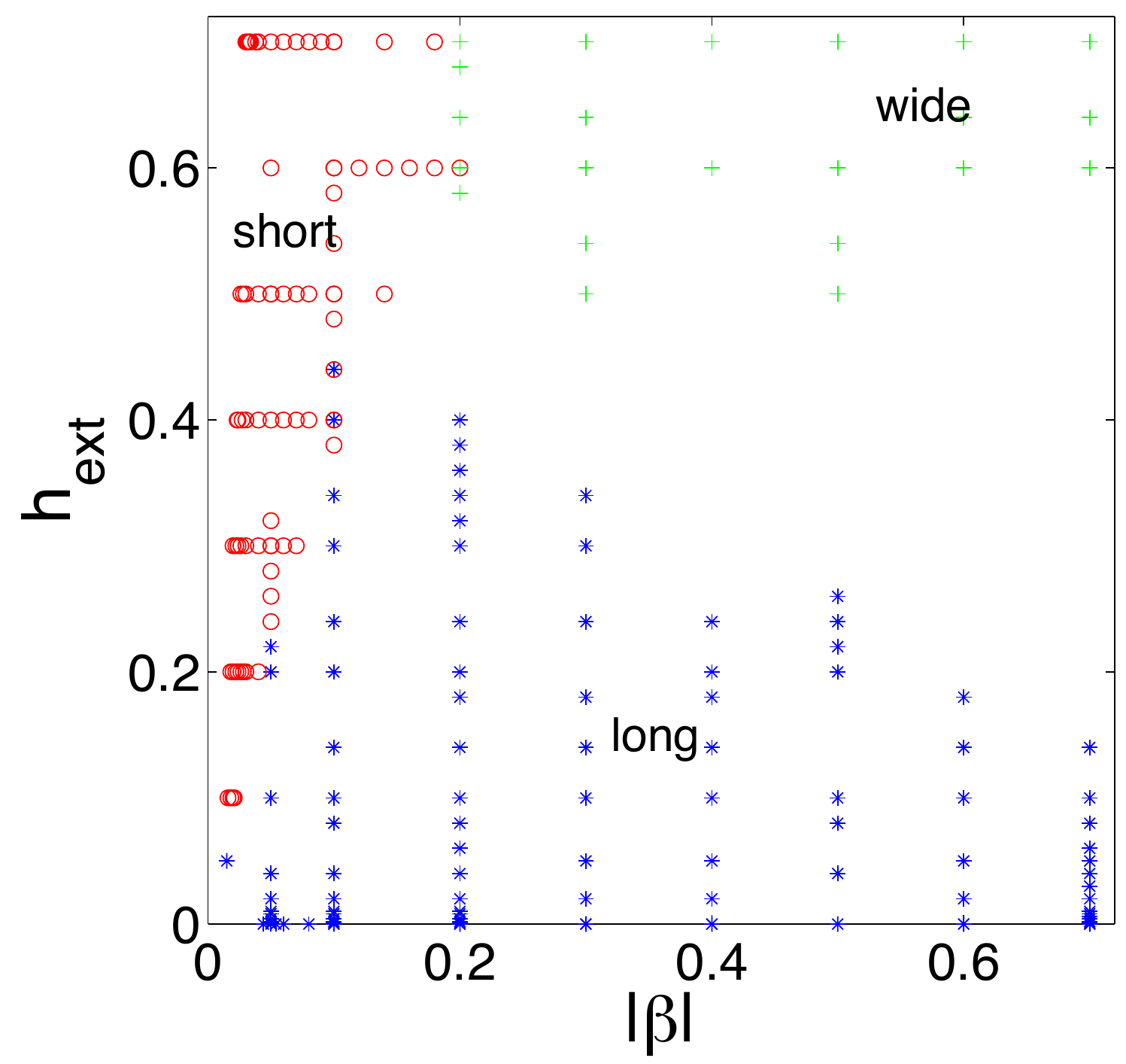} 
   \caption{Numerical simulations have given three types of rotating VA pairs.
   In the $\storque - \hext$ plane we plot blue stars for type I (long) pairs, red circles for type II
   (short) VA pairs, and green crosses for type III (wide) VA pairs.
   No steady-states were found in the regions where no symbols are plotted:
   (a) for very small $\storque$ values and (b) for large $\storque$ and intermediate $\hext$.
   Other parameters as in Eq.~\eqref{eq:parameters}.}
   \label{fig:storque_hext}
\end{figure}

We have explored systematically the range of parameters $0 \leq |\storque| \leq 0.7$
and $0 \leq \hext \leq 0.7$.
We find the three types of VA pairs in steady-state rotation on the $\storque - \hext$ plane
as shown in Fig.~\ref{fig:storque_hext}.
Each symbol corresponds to a numerical simulation which has successfully converged
to a VA pair rotating in a steady-state.
Stars correspond to long VA pairs, circles to short pairs and crosses to wide pairs.
There are small regions in the parameter space where more than one type of VA pairs has been found.
We obtain either one or the other configuration depending on the initial condition used
or on the direction of sweeping the parameter space.
In a significant word of caution we mention that our simulations do not prove that 
the calculated states represent exact steady-states. 
While most of these seem to be exact within our numerical accuracy, in some
others we measure small fluctuations.
Especially the wide pairs for smaller $\hext$ show fluctuations $\sim 1\%-3\%$
in various quantities calculated numerically.

In the areas in Fig.~\ref{fig:storque_hext} where no symbols are plotted 
we have found no steady-state rotating VA pairs.
Instead, the simulations give dynamical states which do include a VA dipole in rotation, 
but the motion is accompanied by
a continuous creation and annihilation of more VA pairs with same polarities.
We will not refer any further to these dynamical states in the present paper.
Simulations and studies of similar states in a related system
have been given in Refs.~\cite{BerkovGorn_PRB2009,BerkovGorn_JPD2008}.

\subsection{Frequency of rotation}

\begin{figure}[t]
   \centering
   \includegraphics[width=2.7in]{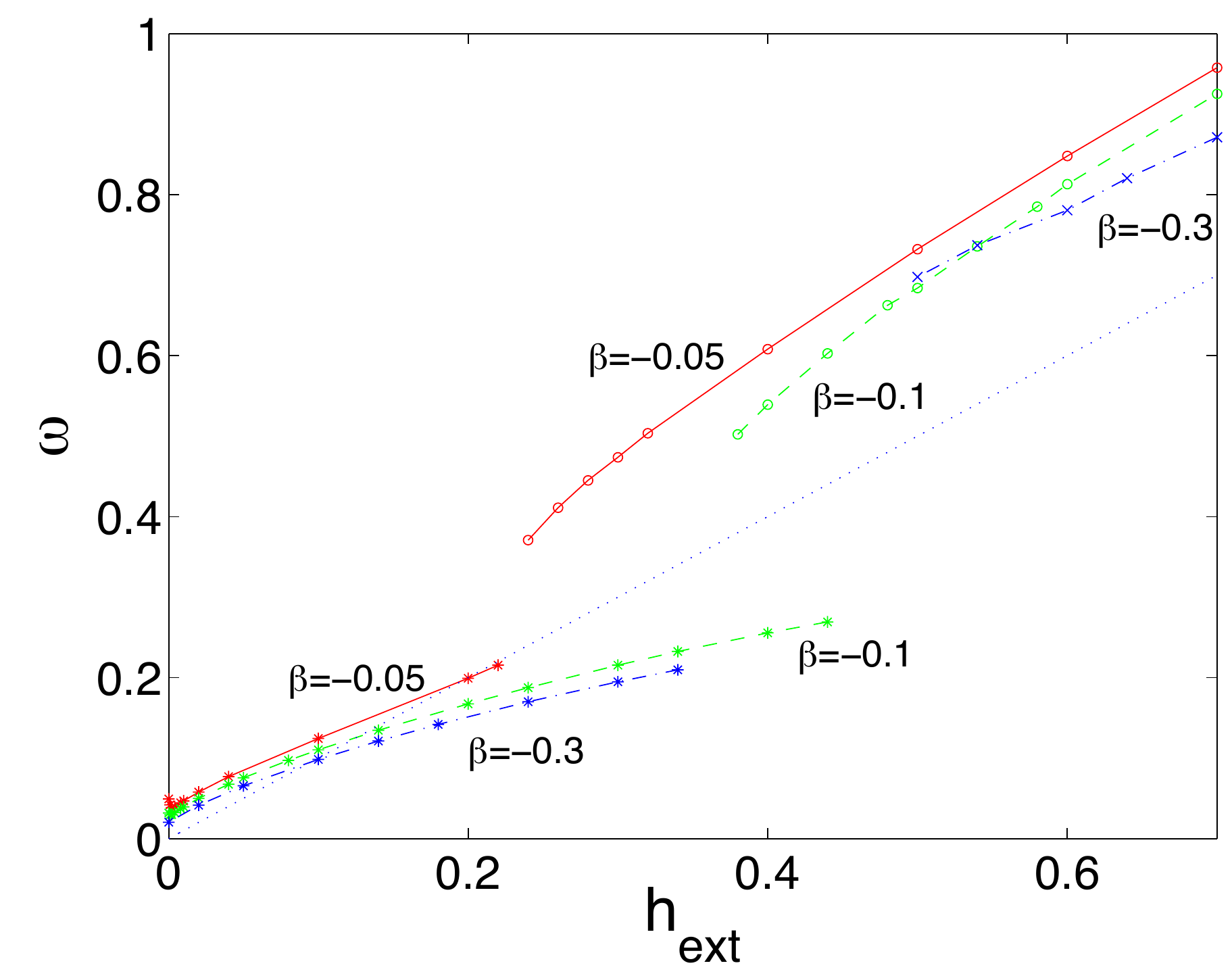} 
   \caption{Angular frequency $\omega$ of rotation (absolute value) as a function 
   of external field $\hext$ for three fixed values
   of the current density $\storque=-0.05$ (solid-red line), $\storque=-0.1$ (dashed-green line),
   and $\storque=-0.3$ (dashed-dotted-blue line).
    Stars denote long VA pairs, circles denote short pairs, and crosses denote wide pairs.
   The frequency $\omega$ goes to a nonzero value for $\hext=0$.
   There is a jump in the frequency when we change from long to short or wide pairs.
   For $\storque=-0.1$ two types of VA pairs coexist for a range of $\hext$ values.
    The dotted line shows $\omega=\hext$ for comparison purposes.
}
   \label{fig:storque_omega}
\end{figure}

For a more detailed presentation of results we fix the spin-torque parameter $\storque$
and vary the field $\hext$.
In Fig.~\ref{fig:storque_omega} we present
the angular frequency $\omega$ of VA pair rotation
for three fixed values of the spin-torque  and
for external field values in the range $0 \leq \hext \leq 0.7$.
For $\storque = -0.1$ and $0 \leq \hext \leq 0.44$ we find long VA pairs and 
for $0.38 \leq \hext \leq 0.7$ we find short VA pairs.
The transition between the two kind of VA pairs is not smooth,  
the magnetic configuration and the angular frequency change significantly.
Furthermore, the two types of VA pairs coexist for the range $0.38 \leq \hext \leq 0.44$.
For $\storque=-0.05$ there is a small gap in the range of $\hext$ where no steady-state rotating
VA pairs are found.
For $\storque=-0.3$ we have long pairs for small $\hext$, wide pairs for large $\hext$
and no steady states for a range of $\hext$ in-between.

VA pairs are certainly stable also for $\hext=0$ and they are rotating at a nonzero $\omega$.
We find $\omega=0.49$ for $\storque=-0.05$ and $\omega=0.31$ for $\storque=-0.1$
and $\omega=0.20$ for $\storque=-0.3$.

\begin{figure}[t]
   \centering
   \includegraphics[width=2.7in]{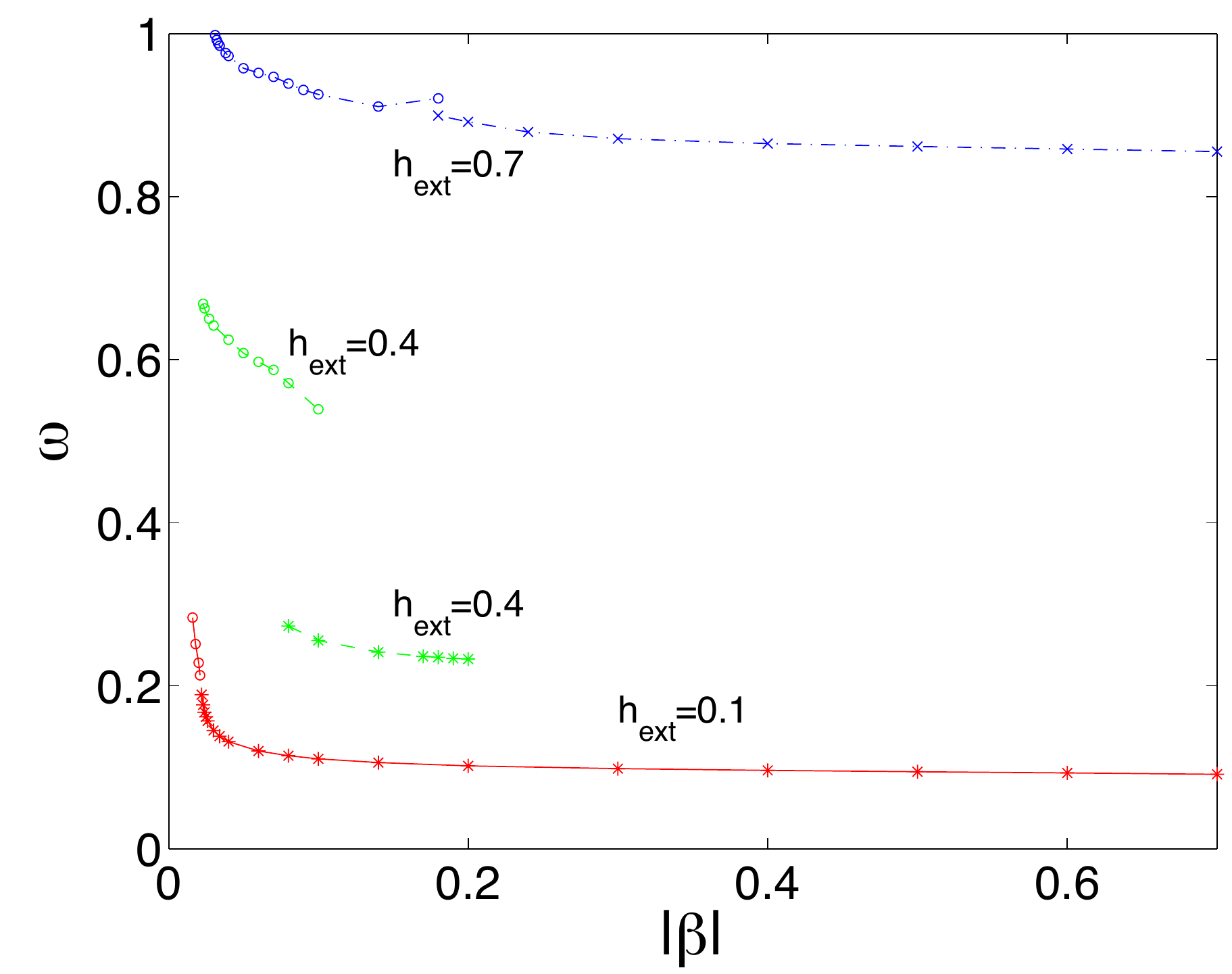} 
   \caption{Angular frequency $\omega$ of rotation as a function of spin-torque $\storque$ 
   for three fixed values of the external field $\hext=0.1$ (solid-red line), $\hext=0.4$ (dashed-green line),
   and $\hext=0.7$ (dashed-dotted-blue line).
   Stars denote long VA pairs, circles denote short pairs, and crosses denote wide pairs.
There is a jump in the frequency when we change from long to short or wide pairs in all case, but it is
 much more pronounced for $\hext=0.4$. No VA pairs are sustained for very small values of $|\storque|$.
 The frequency $\omega$ increases sharply for small $\storque$ and is nearly saturated 
 for large spin-torque values.}
   \label{fig:hext_omega}
\end{figure}

We continue by fixing $\hext=0.1$ and varying $\storque$.
In Fig.~\ref{fig:hext_omega} we present the angular frequency $\omega$ of rotation
for three values of $\hext$ and for spin-torque in the range $0 \leq |\storque| \leq 0.7$.
For $\hext=0.1$ we find long VA pairs for $0.022 \leq |\storque| \leq 0.7$  and 
short VA pairs for $0.016 \leq |\storque| \leq 0.21$.
The transition between the two kinds of VA pairs is not smooth,  
and the angular frequency jumps.
This becomes completely obvious for $\hext=0.4$ where
we have a significant jump to higher frequencies for short VA pairs at $|\storque| \approx 0.1$.
For the values $0.08 \leq |\storque| \leq 0.1$ both long and short pairs are found.
For $\hext=0.7$ we have short VA pairs for smaller $|\storque|$ values and wide pairs
for larger values of the spin-torque.

\begin{figure}[t]
   \centering
   \includegraphics[width=2.7in]{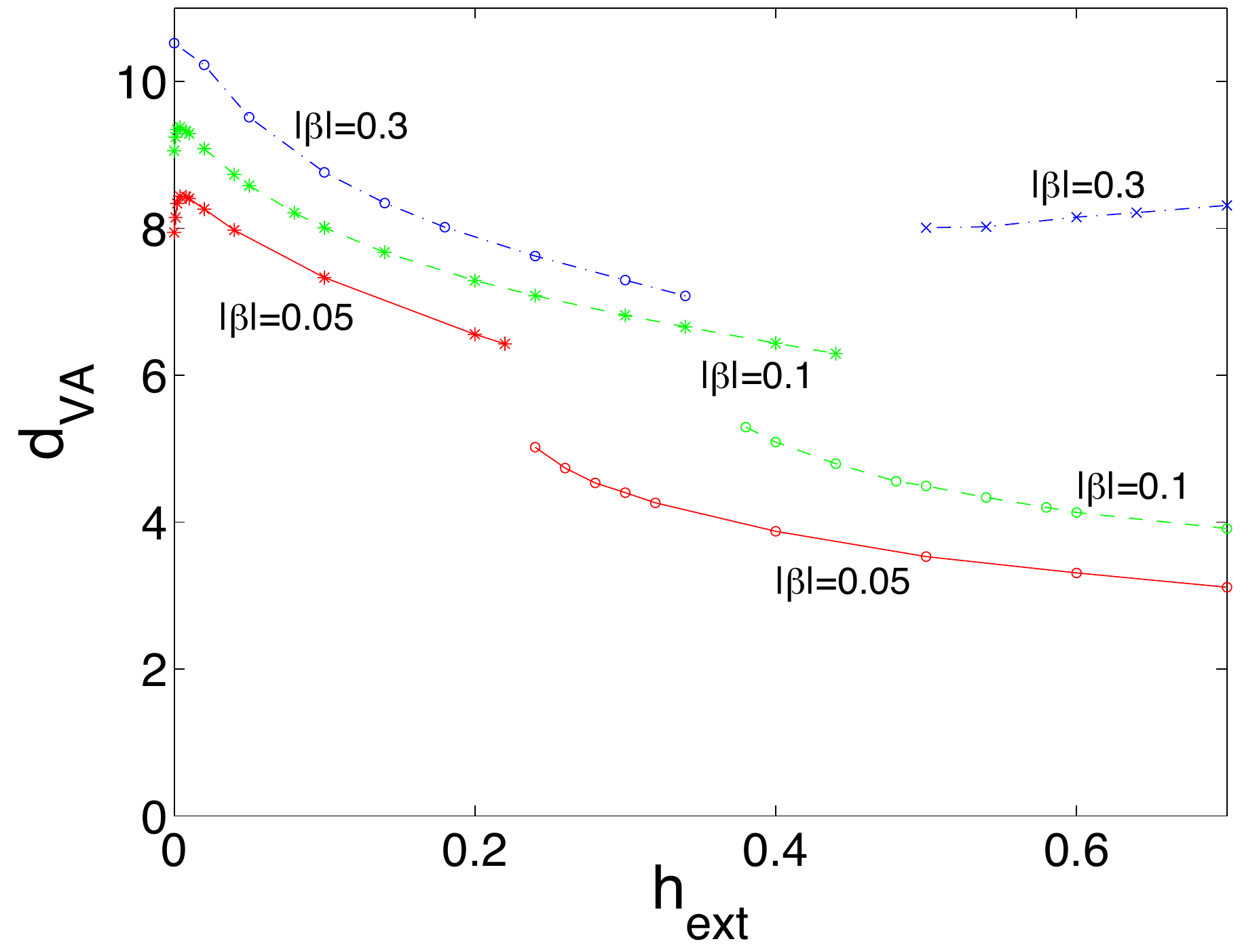} 
   \caption{Separation distance between vortex and antivortex defined in Eq.~\eqref{eq:VAsize}
   as a function of external field $\hext$ for three fixed values
   of the current density $\storque=-0.05$ (solid-red line), $\storque=-0.1$ (dashed-green line),
   and $\storque=-0.3$ (dashed-dotted-blue line).
       Stars denote long VA pairs, circles denote short VA pairs, and crosses denote wide pairs.
  The magnetic field favors the ground state so it acts to shrink the VA pairs, i.e., decrease $\VAsize$.
   The sharp change for $\VAsize$ for very small $\hext < 0.01$ is discussed in Sec.~\ref{subsec:asymptotics}.
   }
   \label{fig:storque_vasize}
\end{figure}

\begin{figure}[t]
   \centering
   \includegraphics[width=2.7in]{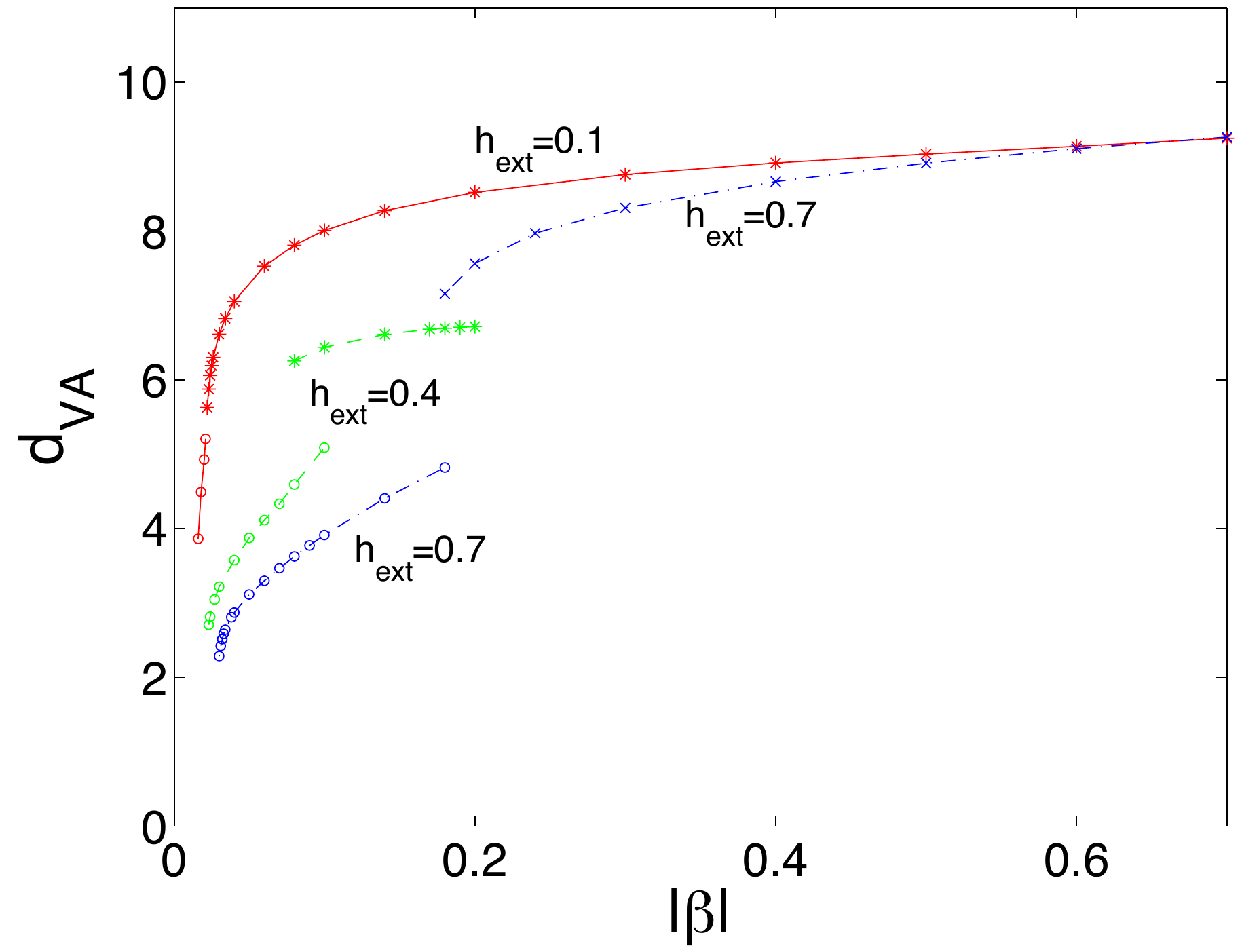} 
   \caption{Separation distance between vortex and antivortex defined in Eq.~\eqref{eq:VAsize}
    as a function of spin-torque $\storque$ 
   for three fixed values of the external field $\hext=0.1$ (solid-red line), $\hext=0.4$ (dashed-green line),
   and $\hext=0.7$ (dashed-dotted-blue line).
   Stars denote long VA pairs, circles denote short VA pairs, and crosses denote wide pairs.
   For very small $\storque$ no VA pairs are sustained. The vortex and antivortex come closer 
   for small $\storque$, and $\VAsize$ saturates for large $\storque$.}
   \label{fig:hext_vasize}
\end{figure}

A quantity of potential interest for experimental work and applications is the distance $\VAsize$
between vortex and antivortex defined in Eq.~\eqref{eq:VAsize}.
Fig.~\ref{fig:hext_vasize} shows $\VAsize$  
as a function of $\hext$ for three values of $\storque$.
Larger values of $\hext$ tend to make the VA pair shrink, as $\hext$ favors the ground state.
Fig.~\ref{fig:storque_vasize} shows $\VAsize$ as a function of $\storque$ for three values 
of external field $\hext$.
The main features are that (i) $\VAsize$ saturates for strong spin torque
and (ii) the vortex and antivortex come very close together for small $\storque$ values
or for large $\hext$.

VA dipoles are sustained even for quite small values of $\storque$.
Smaller $|\storque|$ gives smaller VA pairs and  larger $\omega$.
For very small values of $\storque$ below a certain threshold
(which depends on $\hext$, as can be seen in Fig.~\ref{fig:storque_hext}) 
the simulations show collapse of any initial configuration to the ground state. 
An estimation for the maximum possible $\omega$
is given in Sec.~\ref{subsec:asymptotics}.

It appears straightforward to understand the long VA pairs as a combination of a vortex and an antivortex
which are well-separated but still interacting.
This is in accordance with the overall configuration but also with the detailed features obtained
in the simulations. For example, the anisotropy energy of long VA pairs is roughly twice that
of a single isolated vortex, i.e., $\Ea \approx \pi$.
The short as well as the wide VA pairs are characterised by the fact that their magnetisation 
is roughly similar to their angular momentum,
i.e., $\mu_1 \approx \ell$. This is a feature present in two-meron configurations which are solutions of
the pure exchange model \cite{Gross_NPB1978,Komineas_EPL2012}.
The short and the wide VA pairs thus appear to be related to two-meron configurations, except that, 
unlike merons such as in Eq.~\eqref{eq:VAx},
the VA dipole configuration decays exponentially to the ground state at large distances 
and is thus well-localised.
The wide VA pairs appear to have a large anisotropy energy $\Ea \approx 20$, which is many times
larger than the anisotropy energy of two isolated vortices.

In some cases we can have approximations for $\omega$ directly derived from the virial relation
\eqref{eq:derrick_approx}.
For the case of very small external field the vortex and antivortex are well separated, so we
can use the approximation $\Ea \approx \pi$ and Eq.~\eqref{eq:VAsize}.
Then, the virial relation for long VA pairs, in the case $\hext \ll 1$, gives
\begin{equation}  \label{eq:virial_free}
\omega \approx \frac{2}{\VAsize^2}.
\end{equation}
For long VA pairs at external fields $\hext > 0.1$, we find in simulations $\ell > \mu_1$.
Therefore, the virial relation \eqref{eq:derrick_approx} indicates an angular frequency $\omega < 2/\VAsize^2 + \hext$.

Short VA pairs appear for larger external fields. Since $\mu_1 \approx \ell$
the virial relation gives
\begin{equation}  \label{eq:virial_short}
\omega \approx \frac{\Ea}{\ell} + \hext.
\end{equation}
The first term on the right hand side depends on the precise form of the magnetic configuration.
The approximations given before Eq.~\eqref{eq:virial_free} are not very good any more, however,
one could still use them as a rough guide obtaining a contribution $\sim 1/\VAsize^2$
to the angular frequency.
Approximation \eqref{eq:virial_short} is valid for wide VA pairs, too, 
for which we also have $\mu_1 \approx \ell$.
These VA pairs have large anisotropy energy $\Ea \sim 20$ and also larger angular momentum $\ell$.
Their angular frequency $\omega$ tend to be similar to that for short VA pairs at similar $\hext$ values
as seen in Fig.~\ref{fig:hext_omega}.

\subsection{Asymptotics for large distances}
\label{subsec:asymptotics}

An asymptotic analysis for large distances from the VA pair center is given
in Appendix \ref{app:asymptotics}.
The main conclusion is that
the behavior of the magnetic configuration as $\rho \to \infty$ is given by a system of Bessel equations.
The main features of the solutions depend on the eigenvalues \eqref{eq:mu1mu2},
and their asymptotic behavior \eqref{eq:hankel_asymptotic} on the values of $\lambda_{1,2}$ in Eq.~\eqref{eq:lambda1lambda2}.
The values of these quantities are found
provided the angular frequency $\omega$ is known from the simulations.

A case of special interest is $\mu_1 \geq 0$.
This gives $\lambda_2=0$ in the case of no damping ($\alpha=0$) via Eq.~\eqref{eq:lambda1lambda2}
and thus a non-exponential decay of the magnetic configuration to the uniform state,
as described by the Hankel functions $H_n$ in Eq.~\eqref{eq:hankel_asymptotic}.
If dissipation is present ($\alpha > 0$) then $\lambda_2 > 0$ so we do have
exponential decay, but this is slow, so that unusual behavior could be expected.

Let us see the special case of no external field $\hext=0$, for which we have $\mu_+ \approx \omega^2$.
Such a positive $\mu_1=\mu_+$ means a slow decay for the Hankel functions \eqref{eq:hankel_asymptotic}.
For values of the external field $\hext \sim 0.01$ the numerical data show that 
the eigenvalue $\mu_1=\mu_+$ turns negative and thus $\lambda_2$ acquires larger
values, that is, the decay of the magnetic configuration to the uniform state becomes fast.
The described behavior of the eigenvalue $\mu_1$ explains the great sensitivity,
shown in Fig.~\ref{fig:storque_vasize}, of $\VAsize$ on $\hext$ for very small fields $\hext < 0.01$.
The angular frequency shown in Fig.~\ref{fig:storque_omega} is correspondingly 
very sensitive for small $\hext$, although this is not revealed in the scale of the figure.


Fig.~\ref{fig:hext_omega} shows faster rotation for smaller $\storque$ 
at a fixed $\hext$.
This is due to the vortex and antivortex getting closer to each other.
An increased $\omega$ leads to values for $\mu_1=\mu_+$ closer to zero according to Eq.~\eqref{eq:mu1mu2}.
We have $\mu_+=0$ for angular frequency equal to
\begin{equation}  \label{eq:omega_max}
\omega_{\rm max} = \sqrt{\hext (1+\hext)},
\end{equation}
and we would thus expect the magnetic configuration to tend to delocalise for larger $\omega$ values
($\omega \gtrsim \omega_{\rm max}$).
Indeed, Eq.~\eqref{eq:omega_max} gives a fair approximation to the numerical results
regarding the maximum angular frequency obtained for fixed $\hext \gtrsim 0.1$.
Indeed, no rotation faster than in Eq.~\eqref{eq:omega_max} was obtained in numerical simulations.
We conclude that the slow decay of the configuration is responsible for the unsustainability
of VA dipoles at small spin torque strength. 

An important lesson from the results of the asymptotic analysis is that the rate
of energy dissipation for rotating VA pairs may depend greatly on the features of the magnetisation
configuration. For example, a delocalised configuration would dissipate energy much faster than a localised
configuration. Put it another way, the rate of energy dissipation may not be proportional to 
the value of the dissipation constant $\alpha$.

\section{Uniform spin-current}
\label{sec:uniform}

The results of the previous section showed that when the spin-torque is localised under a nano-aperture
we have a favorable situation for the stabilisation of a VA pair.
In this section we present simulations and show that
there are steady-state rotating VA pairs also in the case that the current flows
through a large area.

We assume in this section an uniform external field \eqref{eq:hext} and 
an uniform spin current polarization \eqref{eq:p}. We will present results for the parameter values
\begin{equation}  \label{eq:parameters1}
\alpha=0.1, \qquad \storque=-0.05 .
\end{equation}
Simulations converge to a steady state rotating VA pair
for external fields $\hext \leq 0.156$.  

\begin{figure}[t]
   \includegraphics[width=3.4in]{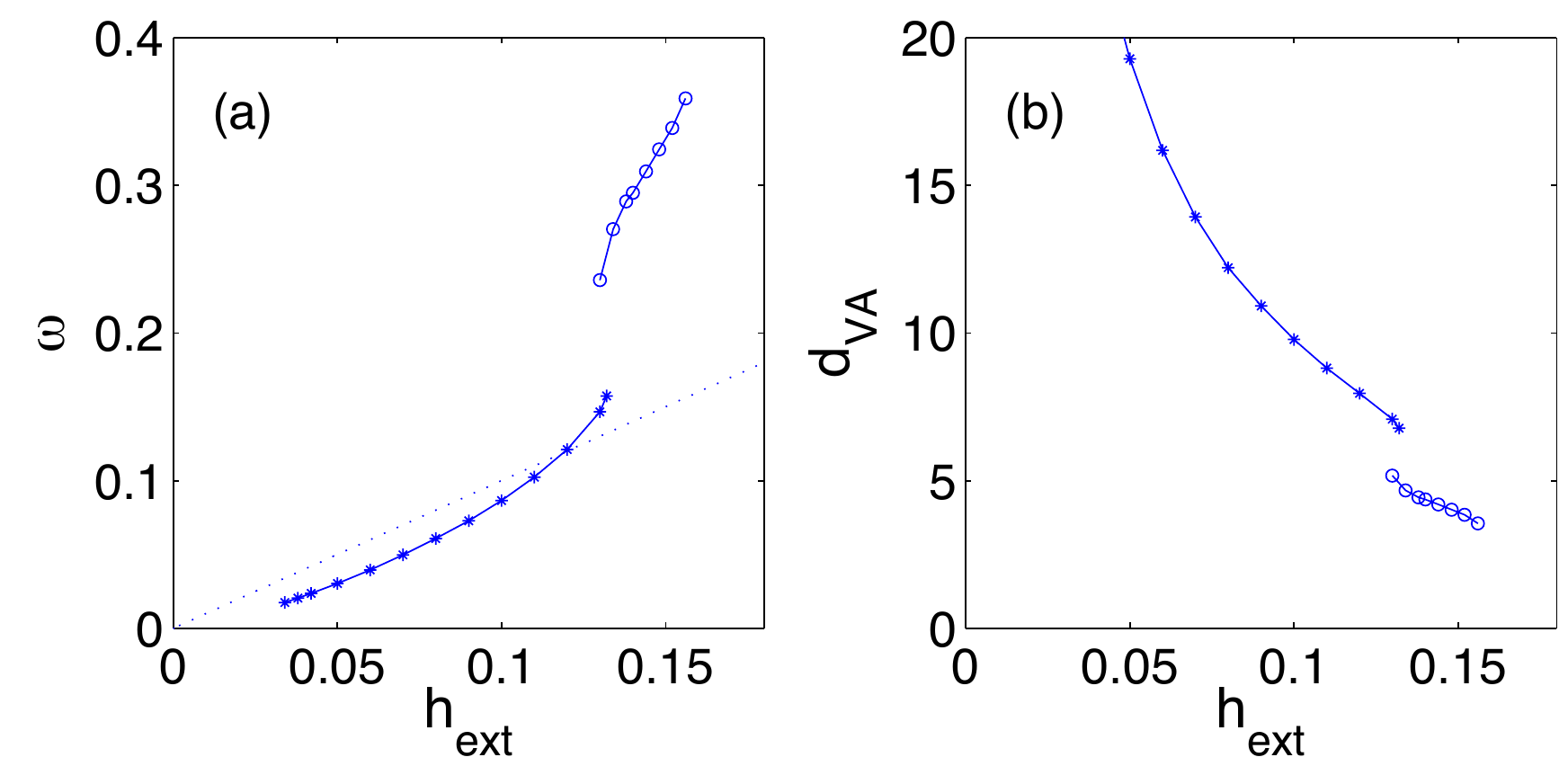}
   \caption{(a) Angular frequency of rotation $\omega$ as a function of the applied field $\hext$.
   Points connected by lines show numerical results for an uniform spin current and 
    parameter set \eqref{eq:parameters1}. We have two separate branches of rotating VA pairs. 
    The lower frequency branch
   (long VA pairs) for $\hext \leq 0.132$ and the upper branch (short VA pairs) for $0.130 \leq \hext \leq 0.156$.
      The dotted line shows $\omega=\hext$ for comparison.
   (b) The separation distance between vortex and antivortex as a function of applied field $\hext$.
   The lower frequency branch corresponds to larger VA separation, while 
   the upper frequency branch corresponds to smaller VA separation.
}
   \label{fig:storque_uniform}
\end{figure}

Fig.~\ref{fig:storque_uniform} shows the rotation frequency $\omega$ and 
the distance $\VAsize$ between the vortices for parameter set \eqref{eq:parameters1} 
and for the range of values of $\hext$ where a steady state was reached.
We find two separate branches of VA pairs. For $\hext \leq 0.132$ we have a branch of 
low frequencies and large VA pair separation, which correspond to long VA pairs.
The branch apparently persists for low values of $\hext$, however, we do not present results 
for $\hext < 0.03$ because the VA separation $\VAsize$ becomes very large 
and our numerical results are then not reliable, 
since the accuracy of the simulations (spatial resolution) for vortices far from the origin is inadequate. 
A separate branch of short VA pairs with higher frequencies and smaller VA pair separation exists 
for $0.130 \leq \hext \leq 0.156$.
For $\hext > 0.156$ we find no steady state rotation (the VA pair is annihilated).
For a narrow range of external fields, $0.130 \leq \hext \leq 0.132$, we have two different steady states and the simulation converges to either a long or a short pair depending on the initial condition.

The Derrick relation \eqref{eq:derrick_approx} is valid here and it is indeed
satisfied with an accuracy better than 1\% in all our numerical simulations.
It provides a guide for the expected frequencies.
For short VA pairs we find $\ell \approx \mu_1$,
so the angular frequency has a significant contribution from the first term on the rhs 
in Eq.~\eqref{eq:derrick_approx} while it is linearly increasing with $\hext$ due to the second term.
For long VA pairs we have  $\mu_1 < \ell$,
so the main contribution to $\omega$ is due to the second term and 
it is approximately proportional to $\hext$.

\section{Conclusions}
\label{sec:conclusions}

Magnetic vortex-antivortex pairs where the vortex and antivortex have opposite polarities
have skyrmion number unity, and the configuration can actually be obtained from the usual
axially symmetric skyrmions by a transformation.
We have studied their dynamics under the influence of spin-polarized current and external magnetic field
using the Landau-Lifshitz-Gilbert-Slonczewski equation.
Both the polarization of the spin current and the magnetic field are in-plane along the same direction.
Their motion is rotational and it is stabilised by the presence of a spin-torque
with in-plane polarisation.
The field contributes directly to the rotation frequency and also indirectly by 
bringing the vortex and antivortex closer together so that they interact stronger.
The rotational dynamics due to both the interaction between vortex and antivortex and the external field
is linked to the skyrmion number.

When an external field is present in the Landau-Lifshitz equation
we would typically expect precession of magnetization around the field.
In the case of the VA dipole we rather have rotation of magnetization.
This surprising result is due to the nonzero skyrmion number of the VA dipole.

The framework developed in the present paper can be applied to spin-transfer
oscillators where the magnetization oscillations are produced due to various external probes,
i.e., due to external magnetic field and spin-torques with various combinations of polarizations.
One possible direction is the oscillations of magnetic bubbles in perpendicular anisotropy
materials, in an analogy to precessing droplets 
\cite{HoeferSilva_PRB2010,MohseniSani_Science2013}.
Also, the effect of perpendicularly oriented polarizers could be analysed
\cite{MoriyamaFinocchio_PRB2012}.

The derivation of most main results is based on the use of the stereographic projection variable \eqref{eq:variableX}.
This allows description of a VA dipole configuration via an axially symmetric anzatz,
and in special cases it leads to the exact description of the dynamics.
Our main results rely upon the form of Eq.~\eqref{eq:llgs} but they do not depend 
on the specific interactions which we included.
The present methods can be adapted to study other cases of counterintuitive
dynamics of vortices, bubbles and other skyrmions in the presence of spin polarized current.

\section*{Aknowledgements}

This work was partially supported by the European Union's FP7-REGPOT-2009-1 project 
``Archimedes Center for Modeling, Analysis and Computation'' (grant agreement n. 245749) 
and by grant KA3011 of the University of Crete.
I gratefully acknowledge discussions with Dimitris Mitsoudis and Dimitris Tsagkarogiannis.

\appendix
\section{Stereographic projection variable}
\label{app:X}

A representation of the magnetisation vector $\bm{m}$ can be given by its stereographic projection 
on a plane.
We define the complex variable
\begin{equation}  \label{eq:variableX}
\Omegax = \frac{m_2+i m_3}{1+m_1},
\end{equation}
which is the sterographic projection of $\bm{m}$ from the point $\bm{m}=(1,0,0)$.
The components of $\bm{m}$ are given as
\begin{equation}  \label{eq:inverseX}
m_1 = \frac{1 - \Omegax\bOmegax}{1 + \Omegax\bOmegax},\quad
m_2 = \frac{\Omegax+\bOmegax}{1 + \Omegax\bOmegax},\quad
m_3 = \frac{1}{i}\,\frac{\Omegax-\bOmegax}{1 + \Omegax\bOmegax},
\end{equation}
where $\bOmegax$ is the complex conjugate of $\Omegax$.
This variable turns out to be particularly useful for studying many properties of the VA dipole.

For the usual conventions adopted in the present paper, 
we have $\bm{m}(\rho\to\infty)=(1,0,0) \Rightarrow \Omegax(\rho\to\infty) = 0$
while $\bm{m}(\rho=0)=(-1,0,0) \Rightarrow  \Omegax(\rho=0) \to\infty$.
At the centers of the vortices we have $\bm{m}= (0,0,\pm 1) \Rightarrow \Omegax = \pm i$.

The Landau-Lifshitz-Gilbert-Slonczewski equation of motion, when we assume external field \eqref{eq:hext}
and spin polarisation \eqref{eq:p}, takes the following form
\begin{equation}  \label{eq:llgsX}
\begin{split}
(i+\alpha)\,\dot{\Omegax} & =  \p_\mu\p_\mu\Omegax - \frac{2\bOmegax}{1+\Omegax\bOmegax}\,\p_\mu\Omegax\p_\mu\Omegax  \\
& -\frac{1}{2}\,\frac{\Omegax-\bOmegax}{1+\Omegax\bOmegax}\,(1+\Omegax^2) -(\hext+i\storque)\,\Omegax,
\end{split}
\end{equation}
where $\mu=1,2$ and summation is implied.

We can further use the complex position $z=x+iy$ on the $xy$-plane,
denote its complex conjugate by $\bz$, so we consider $\Omegax=\Omegax(z,\bz,t)$.
The skyrmion number is given by
\begin{equation}  \label{eq:SkyrmionX}
\Skyrmion = \frac{1}{4\pi} \int \skyrmion\, d^2x,\qquad 
\skyrmion = 4\, \frac{|\p_z\Omegax|^2 - |\p_\bz\Omegax|^2}{(1+\Omegax\bOmegax)^2}.
\end{equation}

Of particular interest here is the form
\begin{equation}  \label{eq:VAx}
\Omegax = \frac{\bar{a}}{\bz},
\end{equation}
where $a$ is a complex constant and $\bar{a}$ its complex conjugate. 
This represents a skyrmion with $\Skyrmion=1$ as can be calculated using \eqref{eq:SkyrmionX}.
Configuration \eqref{eq:VAx} consists of two merons at a distance $2|a|$ apart,
while each of the merons has core size $|a|$ \cite{Gross_NPB1978}.
Note that, at $z = \pm ia$ we have $m_3=\pm 1$, so the two-meron configuration can also be viewed
as a VA dipole where a vortex with negative polarity
is centered at $z=-ia$ and an antivortex with positive polarity at $z=ia$.
The constant $|a|$ gives the vortex core size and the distance between the vortex and the antivortex is $\VAsize=2 |a|$.
More general skyrmion configuration have been studied in Ref.~\cite{Gross_NPB1978}.

\section{A virial relation}
\label{app:virial}

Motivated by the steady state rotating solutions of the exchange model, 
let us assume the existence of similar steady states in the full model \eqref{eq:llgs}.
More precisely, we assume a configuration rigidly rotating at an angular frequency $\omega$, so we have
\begin{equation}  \label{eq:rigid_rotation}
\dot{\bm{m}} = -\omega\, \eln\,x_\lambda \p_\nu\bm{m}.
\end{equation}
This is inserted in Eq.~\eqref{eq:llgs} to obtain virial (integral) relations.
The procedure is developed in Ref.~\cite{Papanicolaou91} and it is applied for the LLG equation.
For an uniform magnetic field \eqref{eq:hext} and a spin-torque term with polarization \eqref{eq:p}
a generalisation of this procedure gives a, so-called, Derrick relation \cite{Komineas_EPL2012}:
\begin{align}  \label{eq:derrick}
  \omega & \left( \ell + \frac{\alpha}{2} \int \epsilon_{\lambda\nu}\, x_\lambda x_\mu d_{\mu\nu}\,d^2x \right) = \nonumber \\
     & - \left( \Ea + \hext\,\mu_1 + \half \int x_\mu \tau_\mu\,d^2x \right),
\end{align}
where
\begin{align}
d_{\mu\nu} & \equiv \p_\mu \bm{m}\cdot \p_\nu \bm{m},   \nonumber \\
\tau_\mu & \equiv -\storque(\bm{m}\times\p_\mu\bm{m})\cdot\bm{p}
\end{align}
and all other symbols are explained in the main body of the paper.
All the integrals are understood to extend over the whole plane.
The Derrick relation \eqref{eq:derrick} is valid for all steady-state rotating solutions.

\section{Asymptotics}
\label{app:asymptotics}

We are interested in the behavior of the magnetization at large distances from the VA pair
as we expect that localisation properties of the configuration could play a significant
role in its stability.
We look in configuration for which $\Omegax(\rho \to \infty) \to 0$, so
we linearise \eqref{eq:llgsX} while it will be convenient to use polar coordinates $(\rho, \phi)$. 
We are interested in steady rotational motion so we assume
 $\dot{\Omegax} = -\omega \p_\phi\Omegax$. The linearised form of  \eqref{eq:llgsX},
 with the latter substitution, is
\begin{equation}  \label{eq:llgsX_linear}
\begin{split}
  \left( \p_\rho^2\Omegax + \frac{\p_\rho\Omegax}{\rho} + \frac{\p_\phi^2\Omegax}{\rho^2} \right) 
  -\frac{1}{2}\,(\Omegax-\bOmegax)  & \\
  -(\hext+i\storque)\,\Omegax + \omega(i+\alpha)\,\p_\phi\Omegax & = 0.
  \end{split}
\end{equation}
We use a Fourier series of the form
\begin{equation}  \label{eq:X_fourier}
\Omegax(\rho,\phi) = \sum_{n=-\infty}^\infty \Omegax_n(\rho)\,e^{in\phi},
\end{equation}
where $\Omegax_n$ are complex functions of $\rho$.
Substituting \eqref{eq:X_fourier} in \eqref{eq:llgsX_linear} we obtain 
a linear system of differential equations for the $\Omegax_n$'s.
These should satisfy Bessel equation
\begin{equation}  \label{eq:bessel_complex}
\left( \p^2_\rho  + \frac{\p_\rho}{\rho} - \frac{n^2}{\rho^2} \right) \Omegax_n = -(\mu_1 + i \mu_2) \Omegax_n,
\end{equation} 
where we have introduced a complex eigenvalue $\mu_1+i\mu_2$.
So we write
\begin{equation}  \label{eq:hankel}
\Omegax_n(\rho) \sim \Omegax_n^0\; H_n\left(\sqrt{\mu_1+i\mu_2}\,\rho \right),
\end{equation}
where $\Omegax_n^0$ are complex constants and $H_n$ are Hankel functions \cite{Watson,AbramowitzStegun}.

Substituting the form \eqref{eq:hankel} in the system of differential equations for $\Omegax_n$ we obtain
an algebraic system which is closed for every pair of unknowns $\Omegax_n^0, \Omegax_{-n}^0$.
The characteristic equation for the system is
\begin{equation}  \label{eq:characteristic}
\begin{vmatrix}
-(M_1 + n\omega) - i M_2  & \frac{1}{2}  \\
 \frac{1}{2}   &  -(M_1 - n \omega) - i M_2 \\
\end{vmatrix} = 0,
\end{equation}
where $M_1 \equiv \mu_1 + \hext +q/2$ and $M_2 \equiv \mu_2 - n\alpha\omega$.
Note that $\storque$ does not appear in this condition as we consider here the case where
the spin-torque term is localised around the origin and is therefore zero at large distances.
We have
\begin{equation}  \label{eq:mu1mu2}
\begin{split}
& \mu_1 = \mu_\pm \equiv -\left( \frac{1}{2} + \hext \right) \pm \sqrt{\frac{1}{4} + (n\omega)^2 }  \\
& \mu_2 = n\alpha\omega.
\end{split}
\end{equation}

We will need to focus on the behavior of the Hankel functions at spatial infinity ($\rho \to \infty$),
which is
\begin{equation}  \label{eq:hankel_asymptotic}
H_n(z) \sim \sqrt{ \frac{2}{\pi z} }\; e^{\pm i\left(z-\frac{n\pi}{2}-\frac{\pi}{4} \right)},  \\
\end{equation}
where $z \equiv (\lambda_1 + i \lambda_2)\,\rho$ with $\lambda_1 + i \lambda_2=\sqrt{\mu_1+i\mu_2}$,
and we should choose the sign for which the exponent is negative.
We require that the decay in Eq.~\eqref{eq:hankel_asymptotic} should be slowest for $n=1$ in view of the
solutions we study in this paper: these are skyrmion solutions, e.g., as in Eq.~\eqref{eq:VAx}.
So we are interested only in the case $n=1$ in Eq.~\eqref{eq:mu1mu2}.
We have
\begin{equation}  \label{eq:lambda1lambda2}
\begin{split} \lambda_1 & = \frac{1}{2}\,  \frac{\alpha\omega}{\lambda_2} \\ 
 \lambda_2 & = \sqrt{ \frac{-\mu_1 + \sqrt{\mu_1^2 + (\alpha\omega)^2}}{2} }.
 \end{split}
\end{equation}
More interesting is the case $\mu_1=\mu_+$ which gives a smaller value for $\lambda_2$
and thus slower decay in Eq.~\eqref{eq:hankel_asymptotic}.
In the case $\lambda_2=0$ the decay of the configuration to the ground state is non-exponential
and then $H_n \sim 1/\sqrt{\rho}$.
The consequences of such behavior are discussed in Sec.~\ref{subsec:asymptotics}.


\end{document}